\documentclass{segabs}
\usepackage{epsfig}
\usepackage{algorithm,algcompatible,amssymb,amsmath}
\usepackage{algpseudocode}
\usepackage{cprotect}
\usepackage{listings}
\usepackage{xcolor}
\usepackage[11pt]{moresize}
\usepackage{hyperref}
\hypersetup{colorlinks,
citecolor=blue,
linkcolor=blue,
urlcolor=blue
}
\begin{document}

\title{Towards automatically building starting models for full-waveform inversion using global optimization methods: A PSO approach via DEAP + Devito}

\renewcommand{\thefootnote}{\fnsymbol{footnote}} 

\author{Oscar F. Mojica$^{1,2}$ and Navjot Kukreja$^{3}$\footnote{1}, $^{1}$SENAI-CIMATEC Supercomputing Center, $^{2}$National Institute of Petroleum Geophysics (INCT-GP) and $^{3}$Imperial College London}

\footer{Example}
\lefthead{Mojica \& Kukreja}
\righthead{Background models for FWI via DEAP + Devito}

\maketitle

\begin{abstract}
In this work, we illustrate an example of estimating the macro-model of
velocities in the subsurface through the use of global optimization methods
(GOMs). The optimization problem is solved using DEAP (Distributed Evolutionary
Algorithms in Python) and Devito, python frameworks for evolutionary and
automated finite difference computations, respectively. We implement a Particle
swarm optimization (PSO) with an ``elitism strategy'' on top of DEAP, leveraging
its transparent, simple and coherent environment for implementing of
evolutionary algorithms (EAs). The high computational effort, due to the huge
number of cost function evaluations (each one demanding a forward modeling step)
required by PSO, is alleviated through the use of Devito as well as through
parallelization with Dask. The combined use of these frameworks yields not only
an efficient way of providing acoustic macro models of the P-wave velocity field
($V_{p}$), but also significantly reduces the amount of human effort in fulfilling
this task.
% the execution of this task. 
%GOMs have already been put in place
%GOMs require much higher computational cost because a huge number of cost function evaluations might be necessary (Sen and Stoffa, 1995), each one requiring a foward modeling.
\end{abstract}
\section{Introduction}
%This is an introduction. \LaTeX\ is a powerful document typesetting system \cite[]{lamport}. An excellent reference is \cite[]{kopka}. The new \textsf{segabs.cls} class complies with the \LaTeX2e\ standard.
Global optimization methods (GOMs) have presented themselves as an alternative
to estimate a starting model for Full Waveform Inversion (FWI) - even when using
real data \citep{galuzzi2018global}. GOMs are an interesting choice since a
proper parameterization technique coupled with sufficient computing power allow
for a reduction of the time and effort required to build an initial
estimate of the velocity model. Simulated annealing (SA) and genetic algorithms
(GA) have been used for a long time to solve geophysical inverse problems, they
(or a variant) have already been used to estimate a starting model for FWI
\citep{sajeva2016estimation,datta2016estimating}. In recent years, however,
particle swarm optimization (PSO) algorithms have rapidly become an attractive
alternative for solving geophysical inverse problems and seem to enjoy an ever
increasing popularity. In line with this trend, we propose to use an
elitism-based PSO to produce a background model that is used as an input for
FWI. The main impediment to PSO's usage as other GOMs is the high computational
cost due to the huge number of cost function evaluations involved. However,
this restriction is limited to cases where the forward modeling is
computationally expensive.
Thanks to advances in high-performance computing, computationally faster forward
modeling is now a reality in many important cases in geophysics, including
seismic methods. Particularly noteworthy in the seismic modeling context is the
emergence of the Devito domain-specific language (DSL) for automated stencil
computations \citep{devito-compiler, devito-api}. Devito provides a Python-based
syntax to easily express Finite differences (FD) approximations of partial
differential equations (PDEs) such as the acoustic wave equation. Although the
use of GOMs to provide an initial estimate for the velocity model comes at a
computational cost, we believe this cost is compensated by the significantly
reduced human effort. Combining different packages in the python ecosystem
allows one to use tested, performance-optimized code instead of reinventing the
wheel. To this end, we use DEAP \citep{DEAP_JMLR2012} which is a framework
written in Python that simplifies the execution of many optimization ideas with
parallelization features. For the parallelization of fitness evaluations over
multiple nodes, we used Dask. DEAP enabled us to integrate Dask and Devito
easily in our application and to incorporate an ``elitism strategy'', which was
used successfully by another GOM in solving the 2D seismic optimization problem
\citep{sajeva2016estimation}. By integrating DEAP, Dask, and Devito, we can
nearly automate the determining of a reliable starting model for FWI, since only
a few parameters have to be set before start the application. We demonstrate the
effectiveness of our approach using synthetic acoustic data of the Marmousi
model.
%Many frameworks already exist, and we do not have to reinvent the wheel but we can learn from each other.
%However, in the era of high performance computing (parallel computing), computationally faster forward modeling is a reality in many important cases in geophysics, including seismic methods. A rapid and continuing rise in the use

%

%In line with this trend, and inspired by the successful application of other GOMs in solving the 2D seismic optimization problem we propose to use an elitism-based PSO built on the DEAP framework (\href{https://github.com/DEAP/deap}{https://github.com/DEAP/deap}) 

%According to this trend, we propose to compare the effectiveness of the elitism-based GA and PSO to solve a 2D seismic optimization problem. The motivation is to validate or refute the widely speculated hypothesis that PSO has the same effectiveness as the GA (same rate of success in finding true global optimal solutions) but with better computational efficiency. The results of this test could prove to be significant for the future development of seismic inversion approaches using GOMs.

\section{Theory}
\textit{Elitist-mutated PSO}: PSO is a population-based
stochastic algorithm and is a member of the broad category of swarm intelligence
techniques based on the metaphor of social interaction.
%PSO is an evolutionary computation technique based on the social behavior metaphor. %developed by Kennedy and Eberhart [4]  %The algorithm simulates the behavior of bird flock flying together in multi dimensional space in search of some optimum place, adjusting their movements and distances for better search. We use the elitist-mutated PSO (EMPSO) described by \cite{nagesh2007multipurpose}. 
The PSO algorithm is initialized with a population of random candidate
solutions, conceptualized as particles. Each particle $\textbf{x}_{i}=
x_{i1},x_{i2},\allowbreak\hdots,x_{iD}$ is assigned a randomized velocity
$\textbf{v}_{i}=v_{i1},v_{i2},\hdots,v_{iD}$ and is iteratively moved through
the $D$-dimensional problem space. It is attracted towards the location of the
best fitness achieved so far by the particle itself $\textbf{p}_{i}=
p_{i1},p_{i2},\hdots,p_{iD}$ and by the location of the best fitness achieved so
far across the whole population $\textbf{g}_{i}= g_{1},g_{2},\hdots,g_{D}$
($gbest$-global version of the algorithm). At iteration $k$, the basic PSO
algorithm \citep{clerc1999swarm} can be described in vector notation as follows:
\begin{equation}
 \textbf{v}_{i}^{k+1}=\chi\left[\textbf{v}_{i}^{k} + c_{1}\textbf{u}_{1}^{k}\otimes(\textbf{p}_{i}^{k}-\textbf{x}_{i}^{k})+ c_{2}\textbf{u}_{2}^{k}\otimes
(\textbf{g}^{k}-\textbf{x}_{i}^{k})\right]
\label{eq:1}
\end{equation}
\begin{equation}
\textbf{x}_{i}^{k+1} =\textbf{x}_{i}^{k} + \textbf{v}_{i}^{k+1},
\label{eq:2}
\end{equation}
In Eq. \ref{eq:1}, $\chi$, $c_{1}$, and $c_{2}$ are the control parameters
called the constriction factor, cognitive parameter, and social parameter,
respectively. The former is a function of $c_{1}$ and $c_{2}$ as reflected in
Eq. \ref{eq:3}.
\begin{equation}
 \chi =\frac{2}{\mid 2 -\varphi - \sqrt{\varphi^{2}-4\varphi}\mid}, \qquad \text{where} \qquad \varphi= c_{1} + c_{2} \ge 4.
 \label{eq:3}
\end{equation}
On the other hand, vectors $\textbf{u}_{1}$ and $\textbf{u}_{2}$ are
$D$-dimensional vectors of uniformly distributed and independent random numbers
in the $[0,1]$ range used to maintain the population diversity ($\otimes$
denotes element-by-element vector multiplication). We used an improved PSO,
named EMPSO \citep{nagesh2007multipurpose}, which introduces an elitist-mutation
strategy into the PSO to improve its performance.
%In the EMPSO algorithm \citep{nagesh2007multipurpose}, an elitist-mutation strategy is introduced into the PSO to improve its performance. 
Pseudocode of the EMPSO algorithm is presented in Fig. \ref{alg_empso}. In 
EMPSO, the elitist-mutation step is computed as follows: first, all particles
are sorted in ascending order based on their fitness function and the index
numbers for the respective particles are obtained; second, the elitist mutation
(EM) is performed on $NM$ worst particles and the respective particle position
vectors are replaced with the new mutated position vectors, whereas the velocity
vectors of these particles are unchanged.
%%%%%%%%%%%%
\renewcommand{\COMMENT}[2][.5\linewidth]{%
  \leavevmode\hfill\makebox[#1][l]{\textcolor{blue}{//~#2}}}
\algnewcommand\algorithmicto{\textbf{to}}
\algnewcommand\RETURN{\State \textbf{return} }
\algrenewcommand\alglinenumber[1]{\small #1:}

\newcommand{\sfunction}[1]{\textsf{\textsc{#1}}}
\algrenewcommand\algorithmicforall{\textbf{foreach}}
\algrenewcommand\algorithmicindent{.8em}

\makeatletter
\newcommand{\algcolor}[2]{%
  \hskip-\ALG@thistlm\colorbox{#1}{\parbox{\dimexpr\linewidth-2\fboxsep}{\hskip\ALG@thistlm\relax #2}}%
}
\newcommand{\algemph}[1]{\algcolor{GreenYellow}{#1}}
\makeatother

%\makeatletter
%\algrenewcommand\ALG@beginalgorithmic{\small}
%\makeatother

\algnewcommand{\WWhile}[1]{\State\algorithmicwhile\ #1\ \algorithmicdo}
\algnewcommand{\EndWWhile}{\unskip\ \algorithmicend\ \algorithmicwhile}
%%%%%%%%%%%%
\begin{figure}[!ht]
  \centering
  \begin{minipage}{.6\linewidth}
  \scriptsize
\begin{algorithmic}[1]
\For{$i \leftarrow 1$ \algorithmicto{} $NM$}
\State $l\gets ASF[i]$ 
\For{$d \leftarrow 1$ \algorithmicto{} $D$}
\If{rand $<$ $p_{em}$}
\State $x_{ld}=g_{d}+0.1\times VR_{d}\times\text{randn}$ 
\Else{} 
\State $x_{ld}=g_{d}$ 
\EndIf
\EndFor
\EndFor
\end{algorithmic}
  \end{minipage}
\cprotect\caption{Pseudo-code of the EMPSO algorithm. $NM$=number of particles
  to be elitist-mutated; $p_{em}$=probability of mutation; $g_{d}=d$-th
  component of global best particle; $ASF$=index of sorted population;
  \verb\rand\=uniformly distributed random number $U(0, 1)$;
  \verb\randn\=Gaussian random number $N(0 , 1)$; and $VR_{d}$=range of decision
  variable $d$.}
\label{alg_empso}
\end{figure} 
\section {Algorithm overview}

\textit{DEAP}: DEAP
(\href{https://github.com/DEAP/deap}{https://github.com/DEAP/deap}) is a
Python-based evolutionary computation framework. We chose it because it provides
many useful features out of the box, and it is current, actively maintained and
well documented. Moreover, DEAP is highly versatile, whereby most central
members of its class hierarchy, such as individuals and operators, are fully
customizable with user defined implementations. We used DEAP's own PSO and added
a function to implement elitism. At first, we create a new population. For each
particle in the population, we calculate the fitness. There are three main
loops. The outer loop is repeated for every generation until the pre-defined
maximum number of generations is reached. At termination, several statistics and
the final population is saved to a log file. The second loop iterates over all
particles. This loop is distributed over multiple nodes in a cluster using Dask.
Each particle's fitness is calculated using Devito following which the particle
is updated as per Eqs. \ref{eq:1} and \ref{eq:2}. The innermost loop is executed
for a prescribed number of particles ($NM$ in Figure \ref{alg_empso}) if the EM
strategy is enabled.
%Within each generation, separate evaluations of particles are distributed on several processors using the DASK library. The fitness function is computed efficently through Devito. 

\textit{Devito}: Devito
(\href{https://github.com/opesci/devito}{https://github.com/opesci/devito}) is a
DSL embedded in Python and is specifically designed for finite differences in
the context of seismic modeling and inversion. It offers a portable framework
for the automated generation of finite-difference code from a symbolic
description of PDEs. It allows the description of arbitrary time-dependent PDEs
as symbolic Python expressions, from which optimized C code implementing a full
time-stepping modeling loop is automatically generated, compiled and executed
from the application environment. The Devito compiler introduces multiple
performance optimizations when it turns the symbolic PDE representation into
stencil code.

\textit{Parallelization}: DEAP provides an easy way to evaluate individuals in a
population on several cores in parallel. The user need merely provide an
implementation of a map function. Here we used the distributed map function from
Dask. Dask is an open source Python library that provides easy interfaces to scale
python code across a large cluster. Although Dask handles arbitrary task graphs,
here we only exploit its map functionality. 
(\href{https://dask.org/}{https://dask.org/}).
%Some more description of DASK required here

Configuration values of the algorithm are specified in a JSON-file. The file
contains forward modeling parameters such as the model size, vertical and
horizontal space sampling or maximum frequency, PSO parameters and variables to
be employed in the model parameterization (See section below). It is a dictionary of dictionaries. 
The values of \verb\cgrid\ dictionary are used to define the number of unknown medium
parameters, whereas optimum values of \verb\pso\ dictionary variables will have
influence on the optimization process and the output. The number of variables we
have to set up is limited to the number of parameters of these two dictionaries.
An example of the JSON file is shown in Figure \ref{json}. The \verb\lambda\ key in the \verb\pso\ dictionary is used to define the restriction range limits $[-\textbf{v}_{max} , \textbf{v}_{max}]$ of the particles velocity according to $\textbf{v}_{max}=(\textbf{x}_{max}-\textbf{x}_{min})\times\lambda$, while the \verb\scale\ key from \verb\cgrid\ dictionary is the number of parameters scaling factor.

\newcommand\JSONnumbervaluestyle{\color{red}}
\newcommand\JSONstringvaluestyle{\color{blue}}
\definecolor{background}{HTML}{EEEEEE}
% switch used as state variable
\newif\ifcolonfoundonthisline

\makeatletter

\lstdefinestyle{json}
{
  showstringspaces    = false,
  keywords            = {false,true},
  alsoletter          = 0123456789.,
  morestring          = [s]{"}{"},
  stringstyle         = \ifcolonfoundonthisline\JSONstringvaluestyle\fi,
  MoreSelectCharTable =%
    \lst@DefSaveDef{`:}\colon@json{\processColon@json},
  basicstyle          = \ttfamily,
  keywordstyle        = \ttfamily\bfseries,
  numbers=left,
  numberstyle=\tiny,
  stepnumber=1,
  numbersep=8pt,
  backgroundcolor=\color{background},
}

% flip the switch if a colon is found in Pmode
\newcommand\processColon@json{%
  \colon@json%
  \ifnum\lst@mode=\lst@Pmode%
    \global\colonfoundonthislinetrue%
  \fi
}

\lst@AddToHook{Output}{%
  \ifcolonfoundonthisline%
    \ifnum\lst@mode=\lst@Pmode%
      \def\lst@thestyle{\JSONnumbervaluestyle}%
    \fi
  \fi
  %override by keyword style if a keyword is detected!
  \lsthk@DetectKeywords% 
}

% reset the switch at the end of line
\lst@AddToHook{EOL}%
  {\global\colonfoundonthislinefalse}

\makeatother
%%%%%%%%%%%%%%%%%%%%%%%%%%%%%%%%%%%%%%%%%
\begin{figure}[!hb]
\centering
\begin{minipage}{.7\linewidth}
\ssmall
\begin{lstlisting}[style=json]
{
 "shotfile":"filtered_shots.file",
 "t0":0.0, 
 "tn":4000.0, 
 "dt":0.61,
 "f0":0.003, 
 "nshots":50, 
 "shape":[369,375], 
 "spacing":[25.0,8.0], 
 "origin":[0.0,0.0], 
 "nbpml":40, 
 "space_order":8,
 "nreceivers":369,
 "first_src_xcoor":175.0,
 "int_btw_shots":175.0,
 "src_depth":0.0,
 "rec_depth":0.0,
 "cgrid":{
          "vstart":1500.0,
          "vend":3500.0,
          "scale":2.5,
          "water_samples":4
         },
 "pso":{
        "lambda":0.5,
        "c1":2.0,
        "c2":2.0,
        "gen":100
       }
}  
\end{lstlisting}
\end{minipage}
\caption{Example JSON Config File. This example contains a set of input parameters for a synthetic case. Such file could change for a real case, but the parameters that control PSO and allow the definition of the number of unknown model
parameters would remain unchanged.}
\label{json}
\end{figure}
%%%%%%%%
%%%%%%%%%%
%A basic PSO is already implemented in DEAP, allowing the
%
\section{Testing approach (EMPSO + local FWI)}
\begin{figure*}[ht]
\centering
     \includegraphics[width=0.65\textwidth]{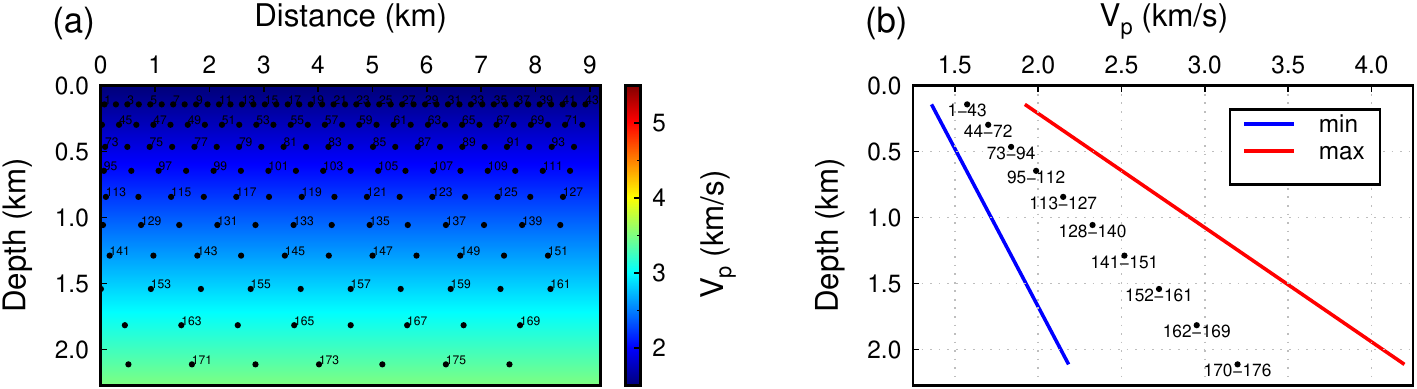}
      \caption{(a) 1D gradient model and the irregular grid nodes. (b) Search range used in the inversions.}
       \label{fig:bounds}
\end{figure*}
%To make the optimization problem computationally feasible, we employ a parameterization technique denominated two-grid (Sajevaet al., 2017), where the velocity model is represented by using a few parameters corresponding to the nodes of a coarse grid with a variable grid spacing, and the forward modeling is carried out over a velocity model mapped on a fine regular grid obtained from the coarse grid by interpolation
In this section, we show the results obtained when the complete workflow (global
+ local search methods) was applied for retrieving a cropped Marmousi model
($285\times369$ samples, with vertical and horizontal space sampling of 8 m and
25 m, respectively). For the forward modeling of the EMPSO algorithm, we use the
FD method (optimized FD stencil code generated with Devito), with an accuracy of
2nd order in time and 8th order in space. The acquisition geometry consisted of
50 sources and 369 receivers, one receiver for each sample on the horizontal
axis. We generated synthetic data using a Ricker wavelet with a maximum
frequency of 17 Hz, and set the sampling and recording times to 0.8 ms and 4
s, respectively. This dataset was then filtered (below 6 Hz), and a Ricker
wavelet with a maximum frequency of 6 Hz was used to compute the modeled
data. To evaluate the misfit, we use the $l_{2}$ norm.
%A trace-by-trace data normalization before the misfit computation was also applied.
Proceeding as in \cite{sajeva2017genetic}, we use a simple 1D $V_{p}$ model
(which together with the water bottom depth constitute the prior information)
with velocities linearly increasing with depth from 1500 to 3500 m/s. This model
is used to center the EMPSO inversion ranges and also to build the irregular
EMPSO grid by following predefined resolution criteria. For full details on such
criteria, see \cite{sajeva2017genetic}. The resulting grid (black dots) and the
linear 1D model are shown in Figure \ref{fig:bounds}-a. This grid has 176 nodes.
These nodes are bilinearly interpolated to the FD grid for the forward-modeling
following what has come to be called a ``two grid strategy''. The ranges for the
$V_{p}$ values during the EMPSO inversion are shown in Figure
\ref{fig:bounds}-b. We defined the minimum and maximum limits for the first and
last level of depth as a percentage of the velocity value of the grid nodes at
these levels. The limits for the intermediate levels of depth are defined by the
lines passing through the maximum and minimum points of the shallower and deeper
levels.

For the EMPSO we used 360 particles, 100 iterations, $NM$=25\%, $P_{em}$=0.3 and
the \textit{gbest} topology \citep{kennedy1999small}. In EMPSO, the EM step
begins from 10th iteration (10\% of the maximum number of iterations) and the
coefficients of cognitive ($c_{1}$) and social($c_2$) acceleration were set to
1.2 and 2.9, respectively. In the optimization process, if the model variables
violate their upper or lower bounds, they are artificially brought back into the
search space, i.e. a box constraint. In the EMPSO inversion, we performed 36,000
model evaluations and the final best-fitting model is used as a starting point
for a local full-waveform inversion.

While Devito could be used in the full workflow (global + local inversions), our
initial work has focused on showing its potential to reduce the time required by
the GOMs solving the wave equations a large number of times. This enables the use
of GOMs in a reasonable amount of execution time. 
The implementation of a complete workflow for FWI built with Devito is a larger
project that we intend to carry out in the near future. The lessons learnt from
this distributed GOM implementation shall be useful in that implementation. 
For this work, instead of using Devito to implement a local FWI, an in-house FWI code is
used to do the job. Our implemented descent-based FWI algorithm uses the
steepest-descent method and a multiscale approach (performing thirty iterations
for each frequency band with maximum frequencies of 4.6, 11.5, 18.4, 25.3, 32.2
and 39 Hz). The line search along the gradient search directions uses the
Barzilai-Borwein (BB) formula for an initial step length
\citep{barzilai1988two}. When required, it applies a backtracking line search
method to update the step length. The forward problem in FWI is formulated in
the time domain and solved using an FD method having an accuracy of 2nd order in
time and 16th order in space (Devito was not used for this), with a time step of 0.8 ms
to ensure stability. The recording time and sampling grid (\textit{dx} and
\textit{dy}) were equal to those used by EMPSO.

% (3 s, $\Delta$x =25 m and $\Delta$z =8 m).

The experiments were run on the \`OG\'UN Supercomputer at SENAI CIMATEC, which
uses an Ethernet interconnection. Each compute node used contains 192 GB of RAM
and two sockets, where each socket has an Intel Xeon Gold 6148 CPU at 2.4 GHz.
\section{Results}

\begin{figure}[h!]
\includegraphics[width=\linewidth]{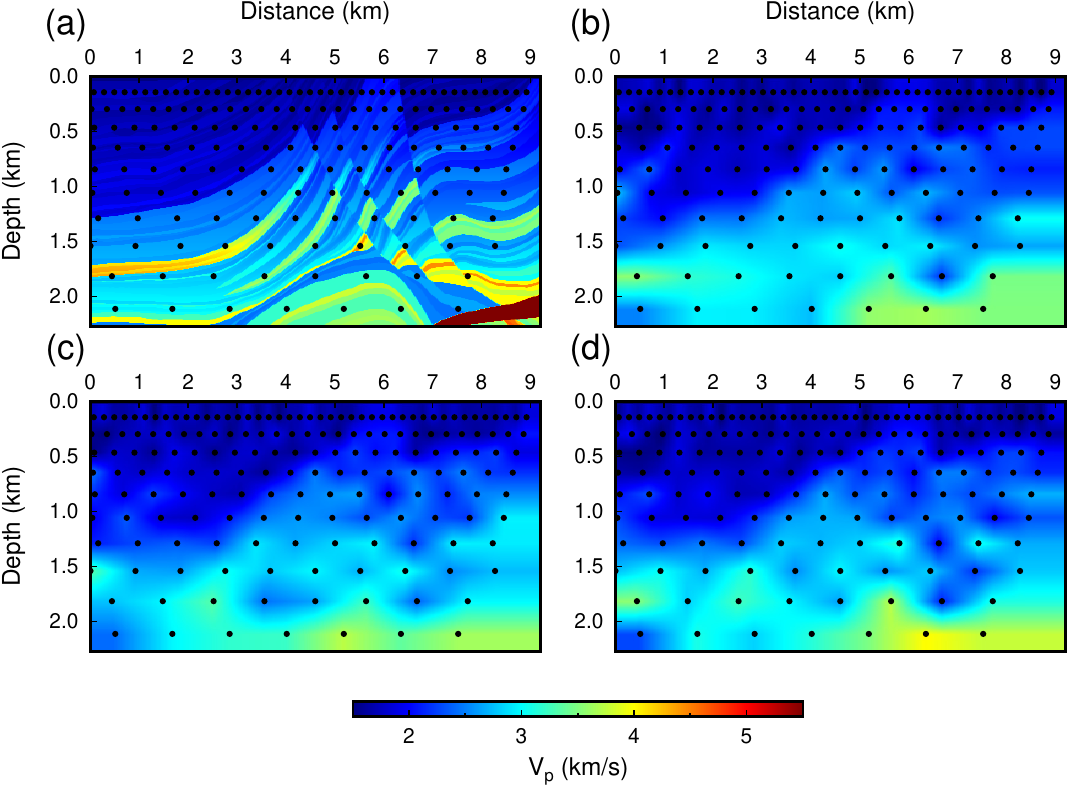}
\caption{(a) Cropped Marmousi model. (b-d) Inversion estimates obtained with EMPSO for three random trials.}
% (the same as in Figure \ref{img1}-a here repeated for ease of comparison).
  \label{img1}
\end{figure}

\begin{figure}[ht!]
\centering
     \includegraphics[width=0.65\linewidth]{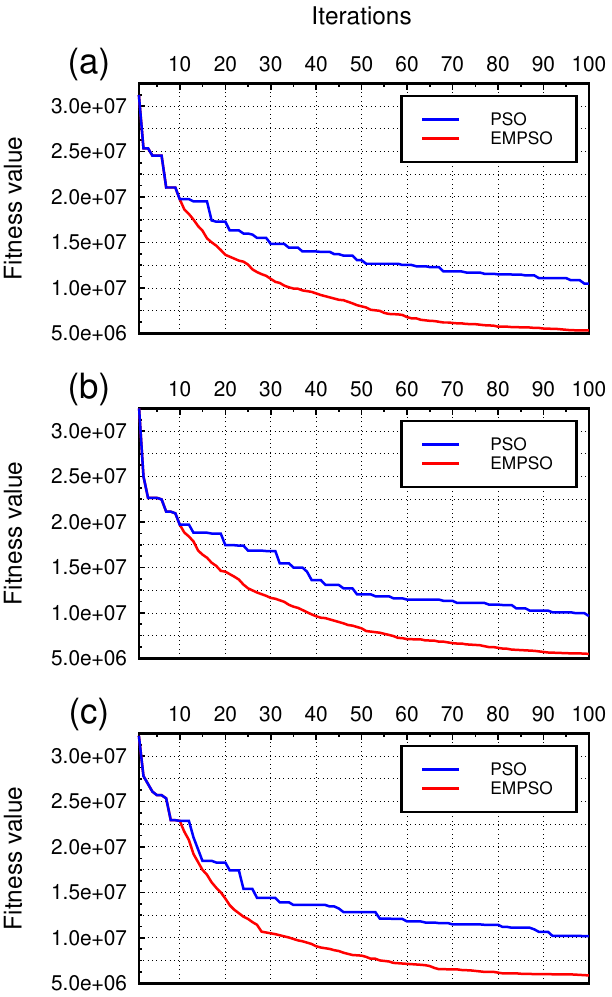}
      \caption{Evolution of the cost function for three different simulations using
conventional PSO and EMPSO. EMPSO clearly outperforms PSO. } % In EMPSO the elistism-mutation is applied after a preselected number of iterations (10\% of maximum number of iterations).}
       \label{fig:error}
\end{figure}

\begin{figure}[h!]
\includegraphics[width=\linewidth]{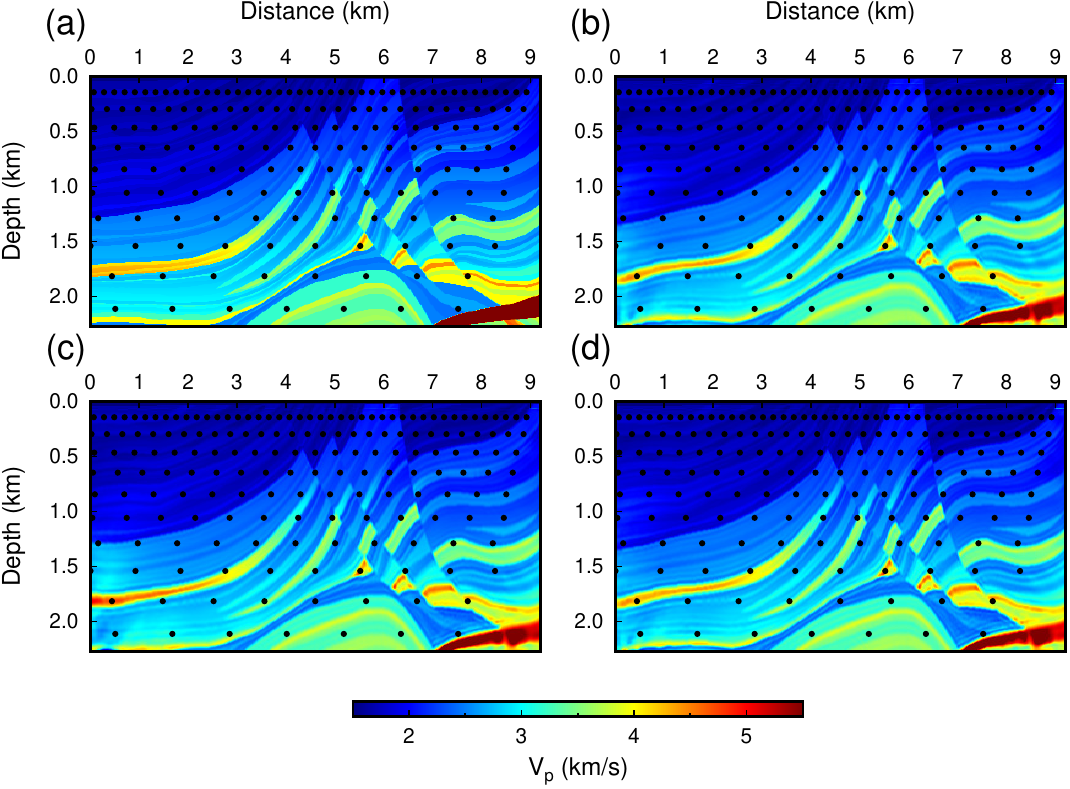}
\caption{(a) Cropped Marmousi model.% (the same as in Figure \ref{img1}-a here repeated for ease of comparison). 
 (b-d) Final models after descent-based FWI from the EMPSO starting models (b-d) of Figure \ref{img1}.}
  \label{img2}
\end{figure}

Figure \ref{img1} ilustrates EMPSO results for three random trials. Figure
\ref{fig:error} shows the comparison of the convergence rate between standard
PSO and EMPSO. For both algorithms, the error gradually decreases over
time. It should be noted, however, that EMPSO gives better fitness values over
different trials than standard PSO. EMPSO does not seem to experience long
periods of stagnation as PSO (apparent from the staircase pattern in fitness
curves). Figure \ref{img2} shows the final models after descent-based FWI using
as starting models the velocity estimates retrieved by EMPSO. The correct
Marmousi model is shown repeatedly in Figures \ref{img1}-a and \ref{img2}-a for
ease of comparison.

In the test scenario outlined in the last section, Devito took less than one
second to generate a single shot gather with 2 threads. It means that the
fitness function evaluation of a particle (50 shots) took approximately one
minute. In the best of cases, with an equal number of available parallel workers
and tasks (fitness evaluations), one iteration would be completed in this time,
which would lead to an extremely fast overall processing time.

In general, the velocity models obtained by FWI using EMPSO outputs as starting
points are very similiar to the true velocity model of Figures \ref{img1}-a and
\ref{img2}-a. Putting it another way, the EMPSO allows the recovery of the low
wavenumber components in the background model to avoid the cycle-skipping
problem, which leads to good final results at the end of the entire workflow.

%As the chosen stopping criterion was the number of iterations/generations neither of the methods stands out over another in computational time. However, it was noted that the final attained value of the fitness in GA trials ($\approx 2\times 10^{6}$) is reached in fewer iterations in EMPSO trials.
%the fitness value for EMPSO trials drops below $2\times 10^{6}$ much faster in comparison with the GA trials. 
%This clearly shows that EMPSO could save computational time using another stopping criterion (i.e., the achievement of apredetermined data misfit value)%, or if the data misfit does not decrease for a certain number of generations).
%and to improveefficiency of the algorithm 
%This, added to the fact that EMPSO  clearly shows could save computational time and to improve efficiency of the algorithm
%\begin{figure}[ht!]
%\includegraphics[width=\linewidth]{error-crop.pdf}
%\caption{xx}
%  \label{img4}
%\end{figure}

\section{Conclusions}

%In order to determine a promising starting model for FWI in due time, we have incorporated a DSL that allows to optimize the execution of successive forward modelings into an Evolutionary Computation Framework, DEAP. automatic generation of highly optimized C code for solving wave equations

In order to determine a promising starting model for FWI, we have incorporated a
DSL, Devito, designed optimally for solving wave equations into DEAP an
Evolutionary Computation Framework, through Dask, a parallelization framework.
With the help of DEAP, we have devised an elitis-based PSO that finds solutions
to the seismic optimization problem. The fitness evaluation (the most
computationally demanding step) is parallelized over a cluster using Dask and
uses Devito to generate highly optimized C code for solving wave equations. The
Devito, Dask, and DEAP mixture offers an optimised and almost automatic
procedure to estimate a starting velocity model for FWI. We test our approach on
a 2D acoustic FWI benchmark problem, namely the Marmousi model. The similarity
between the true and the final models obtained by gradient-based inversions that
started from EMPSO models, makes the proposed approach well suited for the hard
task of finding a good first-guess model for FWI.

%In this paper, we compare the performances of two of the most used GOMs in applied geophysics (GA and PSO) with implemented elitism strategies for estimating acoustic macro models of the $V_{p}$ field using synthetic acoustic data of the Marmousi model. %Both implemented versions of algorithms embody elitism strategies to improve their effectiveness. 
%For the implemented versions, EMPSO yields $V_{p}$ models that provide lower data misfits than those supplied by GA's outputs, although both sets of models reproduce the long-wavelength structures of the Marmousi. Descent-based FWIs results using final GA and EMPSO models are close to the true Marmousi model, demonstrating the ability of both GOMs to yield velocity models suitable as input to descent-based FWI. From our experience, it is observed that EMPSO has faster convergence rate and the best robustness than GA with elitism. EMPSO requires fewer iterations than GA to find the same fitness value, which would have resulted in less CPU time with the proper stopping criterion (supporting the tested hypothesis). However, note that comparisons of GOMs are problem dependent and restricted to the implemented versions.

\section{Acknowledgements}

Computational resources and services used in this work were provided by the High
Performance Computing (HPC) and Research Support Group of SENAI CIMATEC,
Salvador, Brazil. The authors also gratefully acknowledge support from Shell
Brasil through the PSO-FWI project at SENAI CIMATEC and the strategic importance
of the support given by ANP through the R\&D levy regulation. This work was
supported in part by Intel Parallel Computing Centre at Imperial College London
and EPSRC EP/R029423/1.

%The authors also like to thank Dr Justin Lee for his technical support in Genetic Algorithm design.

%\plot*{swarmfinal.pdf}{width=\linewidth}{\label{fig:4} Plot of the $G(w,\lambda)$ curves (eq. \ref{eq11}) for the $28\times50$ (left) and $55\times90$ (right) grids. The $\lambda$ chosen at each iteration is marked by a filled red circle along the iteration number and by a red vertical dashed line. The adopted weight was $w=0.5$.}

\onecolumn

\append{The source of the bibliography}
\verbatiminput{example.bib}

\twocolumn

\bibliographystyle{seg}  % style file is seg.bst
\bibliography{example}

\end{document}